\documentstyle[prl,multicol,aps]{revtex}
\input{epsf}
\tighten

\newcommand{\bx}{{\bf x}}

\begin{document}

\draft

\title{Finite-Temperature Monte Carlo Calculations For
Systems With Fermions}

\author{Shiwei Zhang}
\address{Department of Applied Science 
and Department of Physics, College of William and Mary,
Williamsburg, VA 23187}

\date{\today}
\maketitle

\begin{abstract}
We present a quantum Monte Carlo method which allows calculations on
many-fermion systems at finite temperatures without any sign decay.
This enables simulations of the grand-canonical ensemble at large
system sizes and low temperatures.  Both diagonal and off-diagonal
expectations can be computed straightforwardly.  The sign decay is
eliminated by a constraint on the fermion determinant. The algorithm
is approximate. Tests on the Hubbard model show that accurate results
on the energy and correlation functions can be obtained.
\end{abstract}

%\bigskip


\begin{multicols}{2}
\narrowtext

The quantum Monte Carlo method for simulating grand-canonical
ensembles, originally formulated by Blankenbecler, Scalapino, and
Sugar (BSS)\cite{BSS}, is widely applied in areas spanning
condensed-matter, high-energy, and nuclear physics.  The method allows
essentially exact calculations of finite-temperature equilibrium
properties of interacting fermion systems. It expresses the partition
function as a many-dimension integral over a set of random auxiliary
fields. The many-dimensional integral is then computed by Monte Carlo
(MC) techniques.

As all current fermion quantum Monte Carlo methods, however, the BSS
algorithm suffers from the well-known sign
problem\cite{schmidt84,loh90}.  The integrand of the partition
function is not all positive. Indeed its average sign approaches zero
as the temperature is lowered. As a result, contributions from the
Monte Carlo samples largely cancel. The partition function, which is
given by the difference, becomes a vanishingly small quantity compared
to the MC noise. The computational cost for fixed statistical accuracy
scales exponentially with system size and inverse temperature. While
for many problems the BSS algorithm is the most, sometimes {\em
only\/}, feasible approach, the sign problem has remained completely
uncontrolled in the algorithm.  This has severely limited the
temperatures and sizes accessible, and has prohibited studies of a
variety of interesting problems in correlated systems, particularly
concerning true phase transitions.

In this Letter, we present a finite-temperature method 
which is
free of any decay of the average sign and which retains many of the
advantages of the BSS formalism, thus allowing grand-canonical
calculations at lower temperatures and larger system sizes with
favorable scaling.  Below we first derive a set of {\em exact\/}
constraints on the auxiliary fields which eliminates any negative
contribution to the partition function.  An approximation is then made
to impose these constraints in the MC sampling to control the sign
problem. We develop an algorithm to effectively carry out the MC
sampling under the approximate formalism.  We illustrate the method by
applying it to the one-band Hubbard model.  We show that accurate
results, on both the energy and various correlation functions, can be
obtained with the new method, even with simple forms of the
approximate constraint.

The expectation value of a physical observable $O$ is:
\begin{equation}
  \langle O \rangle = {{\rm Tr} (O e^{-\beta H}) \over 
{\rm Tr} (e^{-\beta H})},
\label{eq:expectO}
\end{equation}
where $\beta=1/kT$ is the inverse temperature.  The chemical potential
term is implicit in the Hamiltonian $H$.  The partition function in
the denominator can be written as
\begin{equation}
Z\equiv {\rm Tr} (e^{-\beta H}) = {\rm Tr} [
e^{-\Delta\tau H}
\cdots e^{-\Delta\tau H} e^{-\Delta\tau H}], 
\label{eq:Z}
\end{equation}
where $\Delta\tau=\beta/L$ and $L$ is the number of ``time slices''
on the right-hand side.

We next write the many-body operator $e^{-\Delta\tau H}$ in terms of
single-particle operators. This is possible for most Hamiltonians or
Euclidean actions of interest.  For example, the Hubbard-Stratanovic
transformation\cite{HS} can be applied for a Hamiltonian $H$ which
contains one- and two-body terms, denoted by $K$ and $V$,
respectively. This transformation replaces the two-body term
$e^{-\Delta\tau V}$ by one-body interactions with a set of random
external fields. Combining the result with the one-body term
$e^{-\Delta\tau K}$, we can write
\begin{equation}
e^{-\Delta\tau H} \simeq \sum_\bx B(\bx),
\label{eq:HS}
\end{equation}
where ${\bf x}$ denotes the random external auxiliary fields
and $B(\bx)$ is a {\em single-particle operator\/}. The sum over all
auxiliary fields recovers the interaction.  For simplicity we have
written the integration over $\bx$ as a discrete sum. We have also
suppressed spin indices, as well as the distribution function of
$\bx$.  The approximation in Eq.~(\ref{eq:HS}) is from the Trotter
error, which is of ${\cal O}(\Delta\tau^2)$ or less.

\begin{figure}
\epsfxsize=3.in
\epsfysize=0.95in
\ \leftline{\epsfbox{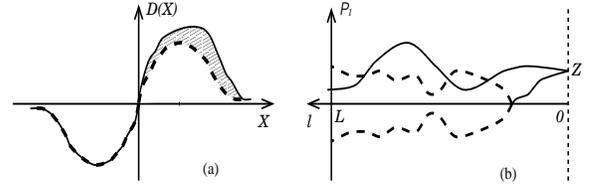}}
\vskip0.05in
\caption{{\em Schematic\/} illustration of the sign problem and the
constraints to control it. Fig.~(a) shows the integrand $D(X)$ of the
partition function $Z$. The $X$-axis represents an abstraction of the
many-dimensional auxiliary-field paths $X$; each point denotes a
collection of $X$'s, e.g., in the sense of a bin in a histogram.  In
standard MC, $|D(X)|$ is sampled, while only the shaded area
contributes.  Fig.~(b) shows ${\cal P}_l$ (Eq.~(\ref{eq:contr})) as a
function of the length of the partial path, $l$, for several paths.
When ${\cal P}_l$ becomes $0$, ensuing paths (dashed lines) cancel.
Only complete paths with ${\cal P}_l>0$ for all $l$ (solid line)
contribute in $Z$; they lead to the shaded area in (a).  }
\label{fig:illu}
\end{figure}

In the standard BSS formalism, Eq.~(\ref{eq:HS}) is substituted into
Eq.~(\ref{eq:Z}).  The trace over fermion degrees of freedom is then
performed {\em analytically\/}\cite{BSS,Hirsch}, which yields
\begin{equation}
{\rm Tr} (e^{-\beta H}) = \sum_X {\rm det}[I+
B(\bx_L) \cdots B(\bx_2) B(\bx_1)],
\label{eq:partition}
\end{equation}
where $X\equiv\{\bx_1,\bx_2,\cdots,\bx_L\}$ denotes a complete path in
auxiliary-field space.  If the size of the single-particle basis ({\em
e.g.\/}, number of spatial lattice sites) is $N$, the single-particle
propagator $B(\bx_l)$ is an $N\times N$ matrix and $I$ is the
corresponding unit matrix.  The fermion determinant, which we will
denote by $D(X)$, can be computed for each $X$.  The sum over all
paths can therefore be evaluated by MC methods.  However, $D(X)$ is
not always positive.  As illustrated in Fig.~\ref{fig:illu}a, the
physical contribution comes from the small difference between the
positive and negative components.  The MC samples of $X$ are drawn
from the probability distribution defined by $|D(X)|$. As $\beta$
increases, $D(X)$ approaches an antisymmetric function and its average
sign vanishes exponentially. The variance in the MC estimate of
Eq.~(\ref{eq:expectO}) thus diverges, and the sign problem occurs.

A main obstacle to understanding and controlling the problem lies in
the implicit and complex nature of the path-integral picture in this
formalism.  To gain insight, we return to the original form of $Z$ in
Eq.~(\ref{eq:Z}). We will use ${\cal B}$ to denote $e^{-\Delta\tau H}$
and imagine the following thought experiment to generate all possible
auxiliary-field paths $X$.  Beginning with ${\rm Tr} [{\cal B} {\cal
B} \cdots {\cal B} {\cal B}]$, we substitute ${\cal B}$ with
Eq.~(\ref{eq:HS}), one at a time from right to left.  After $l$ such
steps, the partition function can be written as
$\sum_{\{\bx_1,\bx_2,\cdots,\bx_l\}} {\cal
P}_l(\{\bx_1,\bx_2,\cdots,\bx_l\},{\cal B})$, where $P_l$ is
\begin{eqnarray}
{\cal P}_l&&(\{\bx_1,\bx_2,\cdots,\bx_l\},{\cal B}) \nonumber\\
&& \equiv 
{\rm Tr} [\;\underbrace{{\cal B}{\cal B} 
\cdots {\cal B}}_{L-l}\:B(\bx_l)\cdots B(\bx_2)B(\bx_1)].
\label{eq:contr}
\end{eqnarray}
As we proceed, we construct paths by including {\em all\/} possible
values of $\bx_l$.  After $L$ steps, all ${\cal B}$'s are replaced and
all complete paths $X$ are generated. Note that, while not the case in
general, the trace in Eq.~(\ref{eq:contr}) can be performed when
$l=L$, which, as expected, gives $D(X)$ of Eq.~(\ref{eq:partition}).

We now examine the procedure more closely, first at $\Delta
\tau \rightarrow 0$, where ${\cal P}_l$ is continuous in $l$,
the length of the partial path.  In particular, we consider the case
when ${\cal P}_l$ becomes zero for a certain partial path
$\{\bx_1,\bx_2,\cdots,\bx_l\}$.  This means that, after the remaining
$L-l$ steps have been finished, the sum over all possible
configurations of $\{\bx_{l+1},\bx_{l+2},\cdots,\bx_L\}$ will simply
reproduce the ${\cal B}$'s in (\ref{eq:contr}), leading to zero by
definition. In other words, any complete path whose first $l$ elements
are $\{\bx_1,\bx_2,\cdots,\bx_l\}$ is ``noise''; the contributions of
such paths cancel in $Z$.  The signature of a noise path is ${\cal
P}_l=0$ for at least one $l$. Since ${\cal P}_0>0$, this
shows that a complete path contributes if and only if the following
$L$ conditions hold:
\begin{equation}
{\cal P}_l(\{\bx_1,\bx_2,\cdots,\bx_l\},{\cal B})>0, \quad l=1,2,\cdots,L.
\label{eq:cpexact}
\end{equation}
If we impose the constraints in Eq. (\ref{eq:cpexact}) in our
procedure to generate the paths, we can eliminate all noise paths
while selecting all contributing paths.  The constraints are
equivalent to having an absorbing boundary at the ${\cal P}_l=0$ axis
in Fig.~1b, thereby making the probability distribution of the
generated complete paths vanish smoothly at the axis.  This boundary
condition (BC) eliminates complete paths that come in contact with
the axis at any point, which cancels out the antisymmetric part of
$D(X)$ in Fig.~1a.  The algorithm remains exact.

At finite $\Delta \tau$, paths are discrete. But the BC is the same
for the underlying continuous paths. To the lowest order, the
constraints in Eq.~(\ref{eq:cpexact}) allow imposition of the BC under
the discrete representation; the contact point (``triple point'' in
Fig.~1b) is approximated by the first $l$ for which ${\cal P}_l<0$.  A
higher order approach, which we use, is to interpolate between this
$l$ and $l-1$, with the probability to terminate at $l-1$ approaching
$1$ smoothly if ${\cal P}_{l-1} \rightarrow 0$\cite{CPMCT=0}. It is
important to note that, in both approaches, the finite-$\Delta \tau$
error vanishes as $\Delta \tau \rightarrow 0$.

${\cal B}$ is not known in practice.  We replace it by a known trial
propagator $B_T$. The constraints now yield approximate results, which
become exact if $B_T$ is exact. If $B_T$ is in the form of a
single-particle propagator, we can analytically evaluate the trace in
Eq.~(\ref{eq:contr}) by making use of the same identity\cite{Hirsch}
that produced Eq.~(\ref{eq:partition}). The constraints in
(\ref{eq:cpexact}) can now be written as:
\begin{equation}
{\cal P}^T_l = {\rm det} [I+(\prod_{m=1}^{L-l}B_T)\:B(\bx_l)\cdots B(\bx_1)]>0
\label{eq:cpcondBT}
\end{equation}
for each $l$ on $1\le l\le L$, where we have introduced the shorthand
${\cal P}^T_l$ for ${\cal P}_l(\{\bx_1,\bx_2,\cdots,\bx_l\},B_T)$. 

The idea of the new method is then to generate MC samples of $X$ which
{\em both\/} satisfy the conditions in (\ref{eq:cpcondBT}) {\em and\/}
are distributed according to $D(X)$\cite{Dnote}.  To realize this
efficiently, we construct
the following algorithm, which builds directly into the
sampling process both the constraints and some knowledge of the
projected future contribution.  In terms of the partial contributions
${\cal P}^T_l$, the fermion determinant $D(X)$ can be written as
\begin{equation}
D(X)={{\cal P}^T_L \over {\cal P}^T_{L-1}}\ {{\cal P}^T_{L-1} \over
{\cal P}^T_{L-2}} \cdots {{\cal P}^T_2 \over {\cal P}^T_1}\ {{\cal
P}^T_1 \over {\cal P}^T_0} \ {\cal P}^T_0.
\label{eq:sampling}
\end{equation}
We construct the path $X$ in $L$ steps, corresponding to stochastic
representations of the $L$ ratios in Eq.~(\ref{eq:sampling}).  We
start from ${\cal P}^T_0$, i.e., $L$ $B_T$'s in place of $B$'s, with
overall weight $1$.  Then, successively from $l=1$ to $L$, we: (a)
pick an $\bx_l$ from the conditional probability density function
$p(\bx_l|\bx_{l-1},\cdots,\bx_2,\bx_1)>0$ defined by $({\cal
P}^T_l/{\cal P}^T_{l-1})/C$ and, (b) multiply the overall weight by
the normalization factor $C \equiv \sum_{\bx_l} {\cal P}^T_l/ {\cal
P}^T_{l-1}$.  The algorithm allows $\bx_l$ to be selected according to
the best estimate of its potential contribution, reflecting the
integrated (i.e., with dashed-line paths in Fig.~(1b) already canceled
out) effect of all subsequent paths from $\bx_l$.  Note that the
probability distribution for $\bx_l$ vanishes smoothly as ${\cal
P}^T_l$ approaches zero, and the constraints are naturally imposed.

We simultaneously propagate an ensemble of paths.  The contribution of
each path $X$ in $Z$ is given by its final weight.  Given $X$, we can
calculate both equal-time and (imaginary) time-dependent correlations
through the single-particle Green's functions\cite{Hirsch}.  The
expectation in Eq.~(\ref{eq:expectO}) is a weighted average over
$X$. The statistical accuracy improves as the procedure is repeated
and more paths are generated.

We mention several technical issues.  (i).~We have chosen a
non-interacting propagator, $e^{-\Delta\tau K}$, as $B_T$. More
general mean-field propagators, including ones with imaginary-time
dependence, can be incorporated straightforwardly.  (ii).~We divide
each step for each path into sub-steps, in which we apply (a) and (b)
to individual components of $\bx_l$. This simplifies $p$ and $C$ (of
the sub-steps)\cite{BSS,CPMCT=0}.  (iii).~As paths are evolved,
products of $B(\bx)$ and $B_T$ must be stabilized against round-off
errors\cite{white}.  (iv).~The weights of paths fluctuate as they are
propagated.  We apply a population control mechanism\cite{pop_cntrl}
to improve efficiency.  A detailed account of these and other
algorithmic issues will be published elsewhere.

The algorithm we have described provides the {\em
finite-temperature\/} counterpart of the {\em ground-state\/}
constrained path Monte Carlo (CPMC) method\cite{CPMCT=0}.  The latter,
which has been applied to study various lattice models, eliminated the
sign decay in $T=0\,$K auxiliary-field calculations by constraining
paths in Slater determinant space with a trial ground-state wave
function $|\psi_T\rangle$ \cite{CPMCT=0,fahy}.  The chief difficulty
in generalizing the concept of a constraining wave function or density
matrix\cite{restr_path} to the finite-temperature formalism is
two-fold: (i) In this formalism, paths do not originate or end at the
same point in Slater determinant space; different paths would thus
require different constraining conditions. Indeed paths do not even
have the same ``dimension''.  (ii) With the analytical evaluation of
the trace, the path-integral picture is implicit and would likely
prevent implementation of such constraints.  The new algorithm
overcame the difficulty. It also provides a unified view of the zero-
and finite-$T$ algorithms. The constraining $|\psi_T\rangle$ in
$T=0\,$K CPMC can be understood in terms of $B_T$ operating on an
initial state.

We now apply the new algorithm to study the one-band Hubbard model.
The model consists of interacting electrons on a square lattice.  The
Hamiltonian $H=K+V$ is given by $K= -t\sum_{\langle ij \rangle \sigma}
(c_{i \sigma}^\dagger c_{j\sigma} + {\rm h.c.})  -\mu \sum_i (n_{i
\uparrow} + n_{i \downarrow})$ and $V= U \sum_i n_{i \uparrow} n_{i
\downarrow}$, where $c_{i \sigma}^\dagger$ creates an electron of spin
$\sigma$ on site $i$, $n_{i\sigma}=c_{i \sigma}^\dagger c_{i\sigma}$
is the electron number operator, and $\langle\;\rangle$ indicates
near-neighbors.  The on-site Coulomb repulsion is $U>0$.  In
connection with high-$T_c$ superconductivity, the Hubbard model has
been the subject of intense theoretical effort for the past
decade. The model provides a good test case, with both its challenging
nature and the availability of certain benchmark data. Quantities of
particular theoretical and experimental interest include the momentum
distribution $n({\bf k})$ and the $d$-wave electron pairing
correlation $P_d({\bf l})$ \cite{dwave_def}.

We study lattices of size $\sqrt N\times \sqrt N$ with periodic
boundary conditions. The desired electron density $\langle n\rangle
\equiv \langle \sum_{i\sigma} n_{i\sigma}\rangle /N$ is achieved by
adjusting $\mu$.  Our trial propagator $B_T$ is $e^{-\Delta\tau K}$
multiplied by $e^{-\Delta\tau \nu_T \sum_{i\sigma} n_{i\sigma}}$,
where $\nu_T$ is a parameter.  The second term in $B_T$ accounts for
$e^{-\Delta\tau V}$ in the sense of restricted Hartree-Fock.

\begin{figure}
\epsfxsize=2.8in
\centerline{\epsfbox{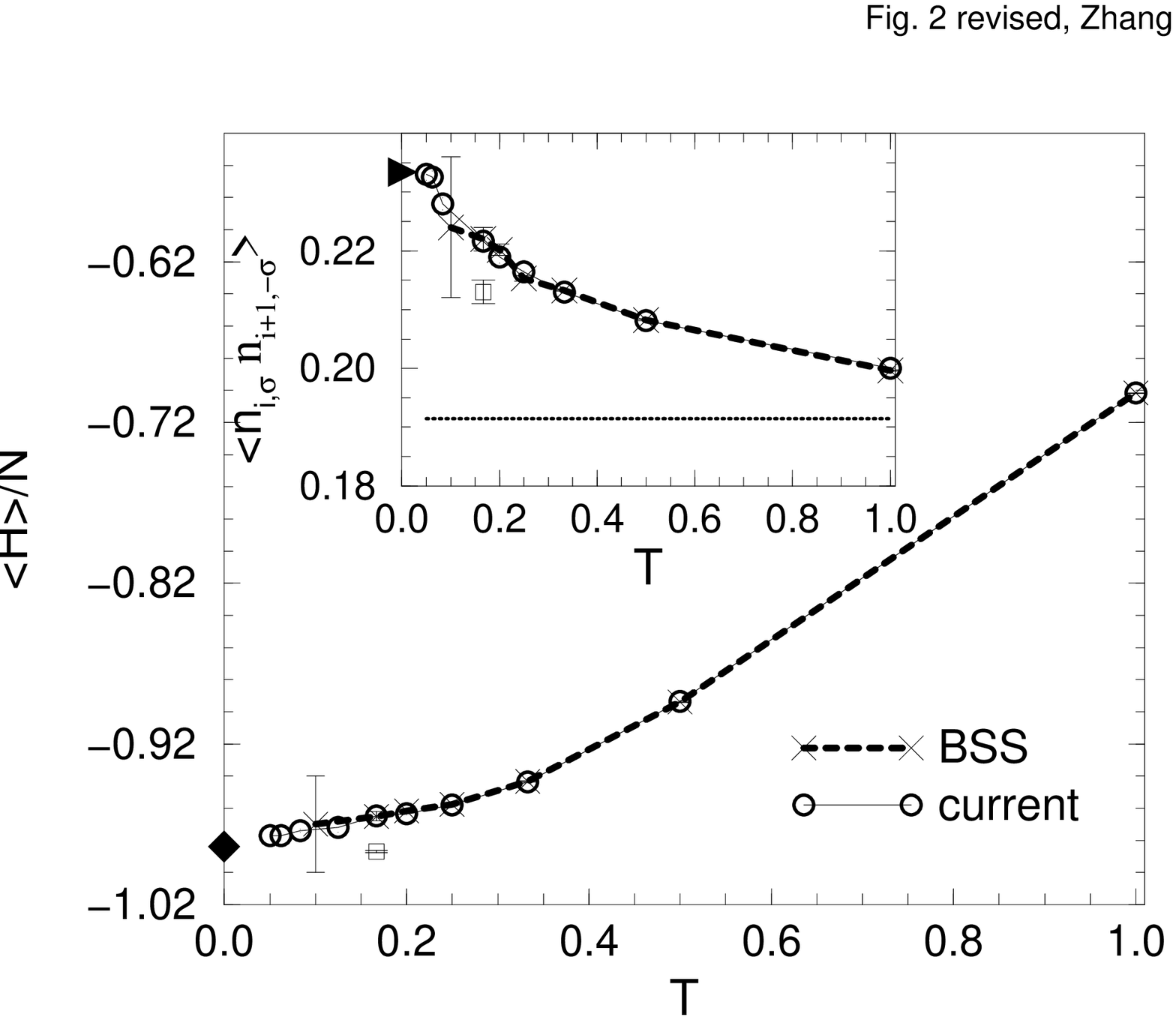}}
\caption{Comparison with available data for a $4\times 4$ system
with $U=4$ and
$\langle n\rangle=0.875$.  The main graph shows the energy. 
The diamond at $T=0$ is from exact diagonalization.
The inset
shows the density-density correlation function between near-neighbor
sites. The algorithm accurately predicts the development of strong
antiferromagnetic correlation as $T$ decreases, despite the use of a
constraining propagator $B_T$ which by itself gives incorrect physics
(flat line). At low $T$, the results converge to that of $T=0\,$K
CPMC (triangle). Error bars in ``current'' are smaller than
symbol size, and are not shown\protect\cite{fig2note}. BSS results are from
Ref.~\protect\cite{RTS_private}.
For comparison, squares at $T=0.1667$ show
BSS results {\em with the sign neglected\/},
which is an uncontrolled approximation\protect\cite{loh90}.
}
\label{fig2}
\end{figure}

In Fig.~\ref{fig2} and in Table I, we show results for a $4\times 4$,
$U=4$ system where the sign problem is the most severe.  This limits
the range of temperatures where accurate calculations can be done
with the standard algorithm.
At $\beta=12$, the average sign in BSS, $\langle s\rangle$,
is projected to be less than $0.01$ from the exponential decay
rate\cite{loh90} and the numbers in Table I;
this $\beta$ is thus not reachable by BSS
with present computing power\cite{Dagotto}.
The system hence presents a
challenging test case for the current algorithm.  At high $T$, our
algorithm gives results in excellent agreement with BSS
results\cite{RTS_private}, which are exact. At low $T$, it reaches
convergence and leads to results consistent with those from
ground-state CPMC and in good agreement with those from $T=0\,$K exact
diagonalization\cite{parola}.

\begin{figure}
\epsfxsize=2.8in
\centerline{\epsfbox{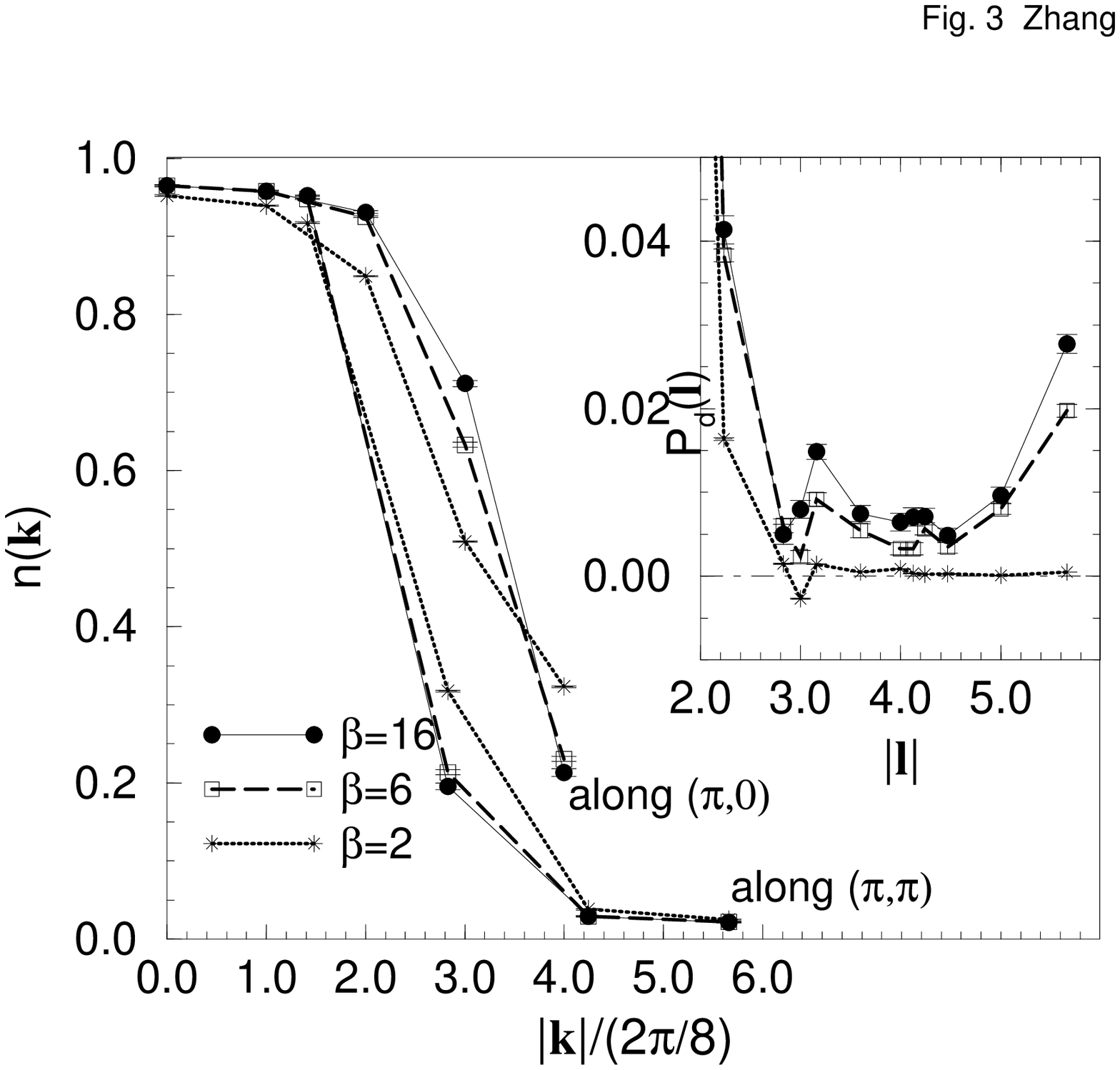}}
\caption{Temperature dependence of the momentum distribution (main
graph) and $d$-wave pairing correlation (inset) for an $8\times 8$
system with $U=4$ and $\langle n\rangle=0.82$.  
(Recall $n({\bf k})$ is the Fourier transform of $G({\bf l})$.)
As temperature ($1/\beta$) lowers, the momentum
distribution, shown along two directions in ${\bf k}$-space, 
becomes more anisotropic, and the
long-range part of the $d$-wave pairing correlation increases.
}
\label{fig3}
\end{figure}

In Fig.~\ref{fig3}, we show new results for an $8\times 8$ lattice.
The electron filling of $\langle n\rangle=0.82$, which is in the
physically relevant region, shows the worst sign problem, with
$\langle s\rangle $ in BSS falling to $\sim 0.1$ at
$\beta=6$\cite{white}.  Accurate and systematic calculations have
therefore not been possible on this system. The new algorithm, on the
other hand, required only modest computing time (about 2 days on a
single processor of an SGI Origin200 workstation for $\beta=16$) to
reach the excellent statistical precision shown in the figure.  As $T$
decreases, the Fermi surface appears to contract along $(\pi,\pi)$,
while bulging along $(\pi,0)$. The $d$-wave electron pairing
correlation at large pair separations {\em increases\/} with
decreasing $T$.  The non-interacting system, however, also shows the
same behavior. In fact, $P_d(|{\bf l}|)$ in the latter is larger
than the corresponding interacting results, consistent with
observations from ground-state CPMC\cite{pairing}.  More systematic
calculations, at different $\langle n\rangle$, $U$, and system size, 
are currently being performed.

In summary, we have presented a quantum MC algorithm which allows
finite-temperature, grand-canonical-ensemble simulations of fermion
systems without any decay of sign.  The method is approximate. We have
shown that accurate results can be obtained with a simple constraining
propagator $B_T$. An improved $B_T$ will lead to improved results, and
the method becomes exact when $B_T$ is exact.  The algorithm makes
possible calculations under the field-theoretical formalism whose
required computer time scales algebraically, rather than
exponentially, with inverse temperature and system size. With the
second-quantized representation, it complements the restricted
path-integral MC method\cite{restr_path}.  The algorithm automatically
accounts for particle permutations and allows easy computations of
both diagonal and off-diagonal expectations, as well as imaginary-time
correlations.  We expect the method and the concept brought forth here
to see many applications, and to significantly enhance the
applicability of quantum simulations in interacting lattice fermion
systems.

I am grateful to R.~T.~Scalettar, F.~Assaad and M.~Enjalran
for providing benchmark data, and to
J.~Carlson for stimulating discussions. I thank D.~L.~Cox,
R.~L.~Sugar, and J.~W.~Wilkins for helpful conversations, and the INT
at the U.~of Washington for hospitality, where part of the work was
done.  This work was supported by the NSF under grant DMR-9734041 and
by an award from Research Corporation.

\newpage

%\end{multicols}

% \begin{figure}
% \caption{}
% \label{}
% \end{figure}

% \begin{table}
% \caption{}
% \label{}
% \end{tabular}
% \end{table}

%\widetext

\begin{table}
\caption{Further comparison of the current method with BSS and exact
diagonalization (ED), on the same system as that of Fig.~2. $G({\bf
l})$ is the average Green's function $\langle c^\dagger_{i+{\bf
l}\sigma}c_{i\sigma}\rangle$, and $P_d({\bf l})$ the $d$-wave pairing
correlation, at separation ${\bf l}=(l_x,l_y)$. The average sign in
BSS is given by $\langle s\rangle$.  In the last row, ED results are
shown for $G$, while ground-state CPMC result for $P_d$; the latter is
{\em not\/} exact. Numbers in parentheses indicate statistical errors
in the last digit.
}
\label{tab1}
\begin{tabular}{c@{\ \ \ \ }l r@{}l r@{}l r@{}l r@{}l r@{}l}
$\beta$ &   & \multicolumn{2}{c}{$\langle s \rangle$}
            & \multicolumn{2}{c}{$G(1,0)$}
            & \multicolumn{2}{c}{$G(2,2)$}     
%            & \multicolumn{2}{c}{$P_{s^\star}(2,1)$}
            & \multicolumn{2}{c}{$P_d(2,1)$}\\ \tableline
$3$ & current &&         & $0.$&$1631(1)$ & $-0.$&$0415(1)$ 
% & $-0.$&$00316(2)$ 
                                                            & $0$.&$0625(2)$\\
    & BSS   &$0.$&$99$ & $0.$&$1631(1)$ & $-0.$&$0418(1)$ 
% & $-0.$&$0488(2)$       
                                                            & $0$.&$0630(3)$\\ \hline 
$6$ & current &&         & $0.$&$1663(3)$ & $-0.$&$0470(4)$ 
%& $-0.$&$0507(3)$ 
                                                            & $0$.&$077(2)$\\
    & BSS   &$0.$&$44$ & $0.$&$1662(2)$ & $-0.$&$0465(2)$ 
%& $-0.$&$0499(3)$ 
                                                  & $0$.&$083(3)$\\ \hline 
$20$ & current & && $0.$&$166(1)$ & $-0.$&$050(1)$   
%& $-0.$&$0498(5)$ 
                                                  & $0$.&$078(2)$\\
$\infty$    & exact    & && $0.$&$167$    & $-0.$&$051$      
%& $-0.$&$0498(4)$ 
                                                  & $0$.&$078(2)$\\
\end{tabular}
\end{table}

\end{multicols}


\begin{references}
\bibitem{Dnote}Note that the constraints in (\ref{eq:cpcondBT}) ensure
$D(X)> 0$, although the reverse is clearly not true.
\bibitem{fig2note}The curves show a small kink at $\beta=6$, because
at $\beta\le 6$, the calculations were done at $\langle n\rangle$
matching that of the BSS calculations, which was larger than
$0.875$ (by $\sim 1$\%), while at $\beta> 6$, they were
adjusted to give $0.875$.
\end{references}
\end{document}